\documentclass[aps,prc,twocolumn]{revtex4}
\usepackage[colorlinks=true,linktocpage=true,linkcolor=blue,citecolor=blue]{hyperref}
\usepackage{graphicx}
\usepackage{amsmath,bbm}
\usepackage{amssymb,bm}
\usepackage{array}

\begin{document}
\title{Drag and Diffusion of Heavy Quarks in a hot and anisotropic 
QCD medium}
\author{P. K. Srivastava\footnote{prasu111@gmail.com}}
\author{Binoy Krishna Patra\footnote{binoyfph@iitr.ac.in}}
\affiliation{Department of Physics, Indian Institute of Technology 
Roorkee, Roorkee~247667, INDIA}
\begin{abstract}
The propagation of heavy quarks (HQs) in a medium was quite often 
modeled by the Fokker-Plank (FP) equation. Since the transport coefficients,
related to drag and diffusion processes are the main ingredients in the
FP equation, the evolution of HQs is thus effectively controlled
by them. At the initial stage of the relativistic heavy ion 
collisions, asymptotic weak-coupling causes the free-streaming 
motions of partons in the beam direction and the 
expansion in transverse directions are almost frozen, hence
an anisotropy in  the momentum space sets in. Since HQs are too
produced in the same time therefore the study of the effect 
of momentum anisotropy on the drag and diffusion 
coefficients becomes advertently desirable. In this article we have 
thus studied the drag and diffusion of HQs in the anisotropic medium 
and found that the presence of 
the anisotropy reduces both drag and diffusion coefficients. In addition, the 
anisotropy introduces an angular dependence to both the drag and diffusion 
coefficients, as a result both coefficients get inflated when the partons
are moving transverse to the direction of anisotropy than parallel to 
the direction of anisotropy. 
\\

 PACS numbers: 12.38.Mh, 12.38.Gc, 25.75.Nq, 24.10.Pa
\end{abstract}

\maketitle 
\section{Introduction}
\noindent
The main outcome of the relativistic heavy ion collision (rHIC) experiments 
is the creation of a deconfined medium of strongly interacting quarks
and gluons, known as quark gluon plasma (QGP). Since the HQs are mainly 
produced at the initial stage of heavy-ion collisions and the thermalization 
time of HQs are of the order of the lifetime of QGP therefore HQs are the 
suitable candidate to probe the QGP.
Due to the large mass of HQ, it was expected from perturbative QCD, the 
nuclear suppression factor, $R_{AA}$ would have been
large, {\em viz.} $R_{AA} \sim$ 0.6 and 0.8 - 0.9 for charm and 
bottom quarks, respectively~\cite{djor,armesto} and 
the elliptic flow (measured as $v_{2}$) of heavy flavoured
hadrons were expected to be smaller than the light hadrons~\cite{armesto}
but the experimental data reveals the opposite 
trend, {\em i.e.} smaller $R_{AA}$ and large 
$v_{2}$~\cite{greco,adare,adler,abelev}. 
Therefore, to circumvent the  contradictory observation, investigation 
for the evolution of HQs with the
proper input and the accompanied energy-loss mechanism in the hot and dense 
QGP becomes essential.\\

For the evolution of HQs in the medium, understanding about the 
energy-loss mechanisms of HQs becomes vital. There are mainly two 
mechanisms for the energy loss of HQs: the first one is
the medium induced gluon radiation (radiation energy loss) and the other 
one is the quasielastic scattering with the background medium partons
(collisional energy loss). Earlier 
it was thought that the medium induced gluon radiation 
is the dominant one but recent studies suggest that
this process is suppressed by the 
large mass of heavy quark, dubbed as ``dead-cone effect''~\cite{dok,lacey}, 
thus the collisional energy loss is then considered 
to be responsible especially at lower energies eg., RHIC energy~\cite{menon}. 
However, at LHC energy the ``dead-cone effect'' is not
so pronounced thus the radiational energy loss become
commensurate again. In brief, the issue is not yet settled 
and one can only say that both mechanisms for
the energy loss are equally important to 
understand the experimental data for both 
$R_{AA}$ and $v_{2}$~\cite{wicks}.\\

Since the evolution of HQs in phase space
can be envisaged as the motion of a nonequilibrated  particle 
in an equilibrium medium therefore the motion of HQs can be
thought as the random motion or the Brownian motion in 
the heat bath of an equilibrated plasma because the mass of HQ
is much smaller than the temperature of the medium. Thus the trajectories
of HQs due to its random motion can be quantified by the Langevin 
dynamics~\cite{patra}, which, however, can also be studied by the 
Fokker-Plank equation~\cite{munshi,moore,vanhees,van,he,alberico} 
in the limit of soft scattering (Landau) approximation. 
Other approaches have also been employed
to study the HQs dynamics at RHIC and LHC 
energies, {\em viz.} relativistic Boltzmann 
transport approach~\cite{zhang,molnar,das4,uphoff}, 
where the Boltzmann equation is 
solved numerically by discretizing the space into a 3-dimensional 
lattice and the collisional integral is modeled by
the stochastic sampling of the collision probability. 
Instead of a constant coupling, the running coupling and the improved 
scattering matrix~\cite{alberico,gossi,cao,ko,surasree} within 
perturbative QCD framework supplemented by hard thermal 
loop (HTL) scheme has been employed to improve upon the collision
integral and thence the drag and diffusion coefficients can be
refined further.  
Since the emergence of hadronic phase is inevitable in 
rHIC therefore some efforts have also been made 
to understand the evolution of heavy flavours in hadronic 
medium~\cite{laine,he1}, which deciphers to subtract
the hadronic contribution from the data to separate the effect of QGP 
alone. Recently~\cite{he1}, authors have shown that even a 
weak coupling of heavy flavour hadrons to the hadronic medium can lead 
to a noticeable contribution to the total elliptic flow. 
The aforesaid discussions are limited to the weak
coupling limit, thus some groups used the complementary 
setup of the gauge-gravity duality~\cite{gubser,solana,horowitz} 
to understand the heavy flavour dynamics at strong coupling limit
in heavy ion collisions. 
In summary, after so many efforts, all models face some difficulties 
to describe both $R_{AA}$ and $v_{2}$ of heavy mesons simultaneously. 

The ultimate aim of the studies on the the drag and 
diffusion coefficients is to determine the transverse 
and the azimuthal momentum distribution of HQs. The fluctuation of the
momentum encoded in the diffusion 
coefficient can be understood in terms of the random forces acting on 
the heavy quarks, which is 
defined by the auto-correlation of the random forces~\cite{laine}.
On the other hand, the drag coefficient, in relaxation-time approximation, 
is related to the kinetic equilibration rate of HQs in 
a thermal medium~\cite{solana}. Asymptotically the 
drag and diffusion coefficients are related by the fluctuation-dissipation 
theorem (FDT): $D/\gamma={\rm{Energy~of~HQ}}~ \times~ 
{\rm temperature}$ (Non-relativistically the FDT relation is
$D/\gamma={\rm{Mass}} \times T$), where $D$ and $\gamma$ are the momentum 
and drag 
coefficients, respectively. One of our aim in this article is to check the
FDT theorem.

Nowadays it is expected that the rHIC collisions at the initial stage may
induce an anisotropy in the momentum space
due to the asymptotic free expansion
of the fireball in the beam direction compared to its transverse direction.
Thus the strongly interacting fluid 
created in rHIC possesses momentum-space anisotropies 
in the local rest frame~\cite{strick,strick1} for
a short duration of time in the initial stage. 
Since the heavy quarks too are produced in the initial stage of 
the collision therefore the momentum-space anisotropies may have 
important implications on heavy quark dynamics. 
This anisotropy subsequently induces the Chromo-Weibal 
instability~\cite{chandra} in the medium, which may significantly 
affect the HQ drag and diffusion coefficients~\cite{chandra}
and facilitates early equilibration of the medium. 
Thus it is desirable to study the effect of momentum-space 
anisotropy on the drag and diffusion coefficients of HQs, which, in 
turn may have significant impact on the 
experimental observables, {\em e.g.} $R_{AA}$ and $v_{2}$~\cite{drag1}. \\

In this article we have thus explored the effect of
momentum anisotropy on the drag and diffusion coefficients 
due to collisional energy loss only when a test 
charm quark evolves in a hot anisotropic QCD medium.
In our calculation we employ the one-loop running coupling constant 
and the Debye mass in the leading and next-to-leading
order to see the effect of the regulator on the $t$-channel matrix 
element, which appears in the collision integral. Our work is thus 
organized as follows: 
First, in subsection II A we revisited the drag 
and diffusion coefficients arises due to collisional energy loss 
alone in an isotropic medium. Here we closely follow 
the kinematics used by Svetitsky~\cite{sve} with the corrected 
matrix elements made in Ref.~\cite{comb}. We then move on to 
an anisotropic medium in subsection II B, where it is found that for weak
anisotropic limit (${\rm{anisotropy~parameter}}, \xi \ll 1$),
both coefficients can be decomposed into the dominant isotopic and the 
sub-leading anisotropic contributions.
Later we demonstrate our results for an isotropic medium in subsection III A 
and understand the salient features
of both coefficients as a function of momentum, temperature etc. 
and its connection with the microscopic properties of HQs evolution
from the point of view of statistical mechanics.
With these understanding in isotropic medium, we then explain the 
numerical results for anisotropic medium in subsection III B.
We have noticed that how the momentum anisotropy affects 
the coefficients and finally transpires to the equilibration rate. 
Finally we conclude in Section IV.
\section{Model Formalism}

\subsection{Isotropic Case}
\noindent
Since the thermalization of HQs is very slow compared to the light quarks
and gluons therefore a description of the motion of
non-equilibrated degrees of freedom in the background
of equilibrated degrees of freedom is required. The appropriate framework 
is provided by the Fokker-Planck equation. Therefore we start with the 
Boltzmann transport equation describing a non-equilibrium statistical 
system as follows:
\begin{equation}
\left(\frac{\partial}{\partial t}+\frac{p}{E}\frac{\partial}{\partial x}+F\frac{\partial}{\partial p}\right)f(x,p,t)=\left(\frac{\partial f}{\partial t}\right)_{coll}.
\label{one}
\end{equation}
For $2\leftrightarrow 2$ interaction the collisional integral 
appearing in the right hand side of above transport 
equation can be written as :
\begin{equation}
\left(\frac{\partial f}{\partial t}\right)_{coll}=\int d^{3}k\left[w(p+k,k)f(p+k)-w(p,k)f(p)\right],
\label{two}
\end{equation}
where $w(p,k)$ is the rate of collision which encodes the change of HQ 
momentum from $p$ to $p-k$ and can be expressed as~\cite{sve}:
\begin{equation}
w(p,k) = g \int\frac{d^{3}q}{(2\pi)^{3}}f(q)v_{\rm{rel}}\sigma_{p,q\rightarrow p-k,q+k},
\label{three}
\end{equation}
Here $f$ is the phase space distribution of the bulk constituents 
, $v_{\rm{rel}}$ is the relative velocity between the two collision partners, 
$\sigma$ represents the cross-section and $g$ is the statistical 
degeneracy of the particles in QCD medium. 

Using the soft-scattering Landau approximation in the collision integral, 
the resulting Fokker-Planck equation (\ref{one}) is cast in the form
\begin{equation}
\frac{\partial f}{\partial t}=\frac{\partial}{\partial p_{i}}\left[A_{i}(p)f+\frac{\partial}{\partial p_{j}}[B_{ij}(p)f]\right],
\label{four}
\end{equation}
where the kernels are defined as :
\begin{equation}
A_{i} = \int d^{3}k~w(p,k)k_{i},
\label{five}
\end{equation}
and
\begin{equation}
B_{ij} = \frac{1}{2}\int d^{3}k~w(p,k)k_{i}k_{j}.
\label{six}
\end{equation}
In low momentum transfer limit ($|p|\rightarrow 0$), kernels are
reduced into
\begin{equation}
A_{i} = \gamma_i p,
\end{equation}
and
\begin{equation}
B_{ij} = D~\delta_{ij}~,
\end{equation}
where $\gamma$ and $D$ are the drag and diffusion coefficient, respectively. 
The generic integral appeared for both the drag and diffusion 
coefficients for HQ in a hot and isotropic  medium
of massless quarks and gluons is given by~\cite{das1} :
\begin{eqnarray}
\langle\langle F_{\rm{iso}}(p)\rangle\rangle &=& \frac{1}{512 \pi^{4}}
\frac{1}{E_p}\int_{0}^{\infty}\frac{q^{2}}{E_q}~dq \nonumber\\
&\times &\int_{-1}^{1}d(cos \chi) \frac{s-M^{2}}{s}~f^{0}(q) \nonumber\\
&\times&\int_{-1}^{1}~d(cos \hat{\theta})\frac{1}{g_{Q}}\sum \overline{|\cal{M}|}^2
 \nonumber\\
&\times& \int_{0}^{2\pi}F({\mathbf p}^{'})~d \hat{\phi}~,
\label{seven}
\end{eqnarray}
where $M$ and $g_Q$ are the masses and degeneracy factors of heavy quarks,
respectively (here we use $M = 1.5$ GeV for charm quark) and $f^{0}(q)$, 
the equilibrium distribution function 
of the massless quarks and gluons is given by
\begin{equation}
f^{0}(q) = \frac{1}{exp\left(\frac{E_{q}}{T}\right)\pm 1},
\label{eight}
\end{equation}
where $\pm$ sign is for quarks and gluons, respectively.\\
Depending on the drag or diffusion, the function, $F(\mathbf p')$ 
in the integral (\ref{seven}) is given by 
\begin{equation}
F(\mathbf{p'}) = \langle\langle 1\rangle\rangle-\frac{\langle\langle{\mathbf p}\cdot {\mathbf p'}\rangle\rangle}{\mathbf{p}^2},
\label{nine}
\end{equation}
for the drag coefficient whereas for the diffusion coefficient
\begin{equation}
F(\mathbf{p}') = \frac{1}{4}\left[\langle\langle \mathbf{p}'^{2}\rangle\rangle -
\frac{\langle\langle({\mathbf p}\cdot {\mathbf p'})^{2}\rangle\rangle} 
{\mathbf{p}^{2}}\right]~,
\label{ten}
\end{equation}
where the dot product, ${\mathbf p}\cdot {\mathbf p'}$ is calculated 
from the expression~\cite{sve}:
\begin{equation}
{\mathbf p}\cdot {\mathbf p'} = E_{p}E_{p}^{'} - \hat{E}_{p}^{2}+
\hat{\mathbf p}^{2}~cos \hat{\theta}
\label{eleven}
\end{equation}
where $E_{p}, ~E_{p^{'}}$ are the energies of incident and scattered 
heavy quarks in the lab frame,
$\hat{E}_{p}$ and $\hat{\mathbf p}$ are the energy and momentum of 
heavy quarks in the CM frame, respectively, and $\hat{\theta}$ is the
CM scattering angle (hereafter the cap represents the variables 
in the CM frame).
The energy, $E_{p}^{'}$ in the laboratory frame is related to the CM frame 
by inverse Lorentz transformation:
\begin{eqnarray}
E_{p}^{'} &=& \hat{\gamma}\left(\hat{E}^{'}+\hat{\mathbf v}\cdot
\hat{\mathbf p}^{'} \right)\nonumber\\
&=& \hat{\gamma}\left[\hat{E}_{p}+|\hat{\mathbf p}|\left(cos\hat{\theta}~\frac{\hat{\mathbf v}\cdot \hat{\mathbf p}}{|\hat{\mathbf p}|}+N~sin\hat{\theta}~sin\hat{\phi}\right)
\right]~,
\label{twelve}
\end{eqnarray}
where the Lorentz factor and velocity in the CM frame are
given by,
\begin{eqnarray}
\hat{\gamma}= \frac{E_{p}+E_{q}}{\sqrt{s}}\\
\hat{\mathbf v} = \frac{{\bf p}+{\bf q}}{E_{p}+E_{q}}
\label{fourteen}
\end{eqnarray}
respectively. The energy, $\hat{E}_{p}$ and the magnitude of the momentum 
$|\hat{\mathbf p}|$ in the CM frame can be written as:
\begin{eqnarray}
|\hat{\mathbf p}|=\frac{s-M^{2}}{2\sqrt{s}}\\
\hat{E}_{p} = \sqrt{{|\hat{\mathbf p}|}^{2}+M^2},
\label{sixteen}
\end{eqnarray}
respectively. Now the dot product $\hat{\mathbf v}\cdot \hat{\mathbf p}$, 
in (\ref{twelve}) can be calculated by the using Lorentz transformation of 
$\hat{\mathbf p}$ from lab frame to CM frame
\begin{equation}
\hat{\mathbf p}= \hat{\gamma}({\mathbf p}-\hat{\mathbf v}E_{p}).
\label{seventeen}
\end{equation}
as 
\begin{equation}
\hat{\mathbf v}\cdot \hat{\mathbf p}=\hat{\gamma}\left({\mathbf p}\cdot\hat{\mathbf v}-|\hat{\mathbf v}|^{2}E_{p}\right)~,
\label{eighteen}
\end{equation}
which can be further simplified as:
\begin{equation}
\hat{\mathbf p}\cdot \hat{\mathbf v}=\hat{\gamma}\left(\frac{|{\mathbf p}|^{2}+{\mathbf p}\cdot{\mathbf q}}{E_{p}+E_{q}}-|\hat{\mathbf v}|^{2}E_{p}\right).
\label{extra}
\end{equation}
The factor, $N$ in Eq. (\ref{twelve}) can be obtained
as~\cite{sve}:
\begin{equation}
N^{2}={|\hat{\mathbf v}|}^{2}-\frac{(\hat{\mathbf p}\cdot \hat{\mathbf v})^{2}}{|\hat{\mathbf p}|^{2}}
\label{ninteen}
\end{equation}
and the Mandelstam variable, $s$ in CM frame is given by
\begin{equation}
s = (E_{p}+E_{q})^{2} - |{\mathbf p}|^{2}-|{\mathbf q}|^{2}-2~|{\mathbf p}||{\mathbf q}|~cos\chi.
\label{twenty}
\end{equation}
where $\chi$ is the angle between ${\bf p}$ and ${\bf q}$.\\ 

In the present work, we consider the collisional energy loss of HQs, 
where the HQs are scattered quasi-elastically with
the partons in QGP medium: $Q (p) + q, \bar q, g (q)\rightarrow 
Q (p') + q, \bar q, g (q')$ (The quantities inside the bracket denotes 
the four momentum of the particle). 
If the HQ is scattered by the quark (anti-quark) then the matrix element 
for the corresponding process is~\cite{sve,comb} :
\begin{equation}
|{\cal{M}}|^{2}_{qc\rightarrow qc} = 256\pi^{2}\alpha_{s}^{2}\left[\frac{(M^{2}-u)^{2}+(s-M^{2})^{2}+2M^{2}t}{(t-\mu^{2})^{2}}\right],
\label{twentyone}
\end{equation}
whereas the matrix element for the gluon scattering is given by~\cite{sve,comb}:
\begin{eqnarray}
|{\cal{M}}|^{2}_{gc\rightarrow gc} &=& \pi^{2}\alpha_{s}^{2}\left[\frac{3072(s-M^{2})(M^{2}-u)}{(t-\mu^{2})^{2}} \right. \nonumber\\
&+&\frac{2048}{3}\frac{(s-M^{2})(M^{2}-u)+2M^{2}(s+M^{2})}{(s-M^{2})^{2}}
\nonumber\\
&+&\frac{2048}{3}\frac{(s-M^{2})(M^{2}-u)+2M^{2}(M^{2}+u)}{(M^{2}-u)}\nonumber\\
&+&768\frac{M^{2}(4M^{2}-t)}{(s-M^{2})(M^{2}-4)}\nonumber\\
&+&768\frac{(s-M^{2})(M^{2}-u)+M^{2}(s-u)}{t(s-M^{2})}\nonumber\\
&-&\left. \frac{256}{3}\frac{(s-M^{2})(M^{2}-u)-M^{2}(s-u)}{t(M^{2}-u)}\right].
\label{twentytwo}
\end{eqnarray}
In the above Eqs. (\ref{twentyone}, \ref{twentytwo}), the $t$ and $u$ 
variables are given by
\begin{eqnarray}
t=2\hat{p}^{2}(cos\hat{\theta}-1) \\
u=2M^{2}-s-t ~,
\end{eqnarray}
and $\mu^{2}$ is the 
regulator, which is needed to shield the infra-red divergences 
arising in the $t$-channel scattering amplitude. In our calculation
we take it as the leading-order Debye mass $m_{D}^{2}$ as~\cite{shuryak}:
\begin{equation}
m_{D}^{2}= T^{2}\left[g^2\left(\frac{N_{c}}{3}+\frac{N_{f}}{6}\right)\right]
\label{twentythree}
\end{equation}
where $g$ is the strong QCD coupling in one-loop. 
\subsection{Anisotropic Case}
Recently it is envisaged that the partonic system generated in
ultra-relativistic heavy-ion collisions at the nascent stage may 
not be necessarily isotropic in the momentum space rather the medium 
exhibits a momentum anisotropy due to the rapid expansion in the longitudinal
direction compared to the transverse 
directions~\cite{dum1,dum2,gribov,mueller,blaiz}.
This motivates us to study the transport coefficients related to drag and 
diffusion processes in such anisotropic medium.

If the anisotropy is small then the anisotropic distribution is
obtained by either stretching or squeezing the isotropic distribution 
along a certain direction, thereby preserving a 
cylindrical symmetry in momentum space. In particular,
the anisotropic distribution relevant for relativistic heavy ion
collision can be approximated
by removing particles with the large momentum component along 
the direction of anisotropy, ${\bf n}$ as~\cite{roma1,roma2} :
\begin{equation}
f_{\rm{aniso}} ({\mathbf q})=f_{iso}\left(\sqrt{q^{2}+\xi({\mathbf q}
\cdot {\mathbf n})^{2}}\right),
\label{twentyfour}
\end{equation}
where $f_{iso}$ is an arbitrary isotropic distribution 
function and $\xi$ is the anisotropic parameter and is  
generically defined as:
\begin{equation}
\xi=\frac{\langle {\bf q}_{T}^{2}\rangle}{2\langle q_{L}^{2}\rangle}-1,
\label{twentyfive}
\end{equation}
where $q_{L}={\mathbf q}.{\mathbf n}$ and ${\mathbf q}_{T}
={\mathbf q}-{\mathbf n}({\mathbf q}.{\mathbf n})$ are the 
components of momentum parallel and perpendicular to  
$ {\mathbf n} $, respectively. There have been significant
advancement in the dynamical models used to simulate plasma evolution 
having momentum-space anisotropies~\cite{Martinez:2010sd-12tu,
Martinez:PRC852012,Ryblewski:2010bs,Ryblewski:2012rr,Florkowski:2010cf}.
One of us have studied the effects of momentum anisotropy on 
the quarkonia states by the leading-anisotropic
correction to the resummed gluon propagator~\cite{lata:PRD2013,lata:PRD2014}
which subsequently affects the suppression of quarkonium production
at RHIC and LHC. Recently we have investigated the effect of momentum 
anisotropy on one the transport coefficients, {\em namely} the electrical 
conductivity~\cite{pks2}.

If the distribution function is nearly an ideal gas distribution and 
the anisotropy, $\xi$ is small then $\xi$ can be related to the 
shear viscosity of the medium via the one-dimensional Bjorken expansion in 
the Navier-Stokes limit~\cite{asa}::
\begin{equation}
\xi=\frac{10}{T\tau}\frac{\eta}{s},
\label{twentysix}
\end{equation}
For an expanding system, non-vanishing viscosity implies 
the finite relaxation time in the momentum space, hence 
an anisotropy of the particle momenta does appear inherently, 
{\em for example} for the ratio, $\eta/s$ in the range, 0.1 - 0.3 
and $\tau T = 1 - 3$, one finds $\xi$ tentatively as $\xi=1$. \\

As we have explained, hot QCD medium due to 
expansion and non zero viscosity, exhibits a local 
anisotropy in momentum space, therefore the quark distribution 
function in the anisotropic medium can be approximated
for a baryonless medium ($\mu_B=0$):
\begin{equation}
f_{\rm{aniso}} ({\bf q};T)=\frac{1}{e^{(\sqrt{{\bf q}^{2}
 + \xi({\bf q}.{\bf n})^{2}+ m^{2}})/T}+1}.
 \label{twentyseven}
 \end{equation}
For weakly anisotropic systems ($\xi<<1$), one can expand the 
distribution function and keep the leading term in $\xi$ only:
\begin{eqnarray}
f_{\rm{aniso}} ({\bf q};T)&=&\frac{1}{e^{E_{q}/T}+1}-\frac{\xi}{2E_{q}T}({\bf {q\cdot n}})^{2}\frac{e^{E_{q}/T}
}{(e^{E_{q}/T}+1)^{2}},\nonumber\\
&=&f^{0}(q)-\frac{\xi}{2E_{q}T}{({\bf q}\cdot{\bf n})}^{2}{f^0}^{2} e^{E_{q}/T},
\label{anisodist}
\end{eqnarray}
where ${\bf q}$ $\equiv$ $ (q\sin\chi\cos\Phi, 
q\sin\chi\sin\Phi, q\cos\chi) $ and ${\bf n}$ $\equiv$ 
$(\sin\beta, 0, \cos\beta)$ or $(0,\sin\beta,\cos\beta)$. $ \beta $ 
is the angle 
between ${\bf q}$ and ${\bf n}$. 

Therefore the drag and diffusion coefficients in 
a weakly anisotropic medium may be obtained by replacing the
phase-space distribution in the anisotropic medium (\ref{anisodist}) 
\begin{eqnarray}
\langle\langle F(p)\rangle\rangle &=& \frac{1}{1024 \pi^{5}}\frac{1}{E_p}\int_{0}^{\infty}\int_{0}^{\pi}\int_{0}^{2\pi} \\ \nonumber
&\times&\frac{q^{2}}{E_q}~sin\chi~dq~d\chi~d\Phi\frac{\omega^{1/2}}{s}~f_{aniso}(q)\\ \nonumber
&\times&\int_{0}^{\infty}sin\hat{\theta}~d\hat{\theta}\frac{1}{g_{Q}}\sum \overline{|M|}^2\\ \nonumber
&\times&\int_{0}^{2\pi}F({\mathbf p}^{'})~d\hat{\phi}.
\label{twentynine}
\end{eqnarray}
Since $f_{\rm{aniso}}$ has two part : isotropic and correction 
due to momentum anisotropy. The resulting expression for drag 
and diffusion has two parts. The isotropic part is similar to 
Eq.(\ref{seven}). The expression for anisotropic part is as follows :
\begin{eqnarray}
\langle\langle F_{\rm{aniso}}(p)\rangle\rangle &=& -\frac{1}{1024 \pi^{5}}\frac{1}{E_p}\int_{0}^{\infty}\int_{0}^{\pi}\int_{0}^{2\pi}\\ \nonumber
&\times&\frac{q^{2}}{E_q}~sin\chi~dq~d\chi~d\Phi \frac{\xi}{2E_{q}T}({\mathbf q}\cdot{\mathbf n}) \\ \nonumber
&\times&{f^{0}}^{2}(q)e^{E_{q}/T}\frac{\omega^{1/2}}{s}\\ \nonumber
&\times&\int_{-1}^{1}d(cos\hat{\theta})\frac{1}{g_{Q}}\sum \overline{|M|}^2\\ \nonumber
&\times&\int_{0}^{2\pi}F({\mathbf p}^{'})~d\hat{\phi}.
\label{thirty}
\end{eqnarray}
Now using the definition of ${\mathbf q}$ and ${\mathbf n}$, 
one can get the expression for $({\mathbf q}\cdot{\mathbf n})^{2}$ as follows :
\begin{eqnarray}
({\mathbf q}\cdot{\mathbf n})^{2}&\equiv& q^{2}sin^{2}\chi cos^{2}\Phi sin^{2}\beta\\ \nonumber
&+&q^{2}cos^{2}\chi cos^{2}\beta +2q^{2}sin\chi cos\chi sin\beta cos\beta cos\Phi.
\label{thirtyone}
\end{eqnarray}
Putting this value in Eq. (\ref{thirty}) and integrating over $\Phi$, 
we can get the modified expression as follows for the 
anisotropic correction part :
\begin{eqnarray}
\langle\langle F_{\rm{aniso}}(p)\rangle\rangle &=& -\frac{\xi}{1024 \pi^{5}}\frac{1}{2E_{p}T}\int_{0}^{\infty}\int_{-1}^{1}\frac{q^{4}~dq~d\chi}{E_{q}^{2}}\\ \nonumber
&\times&\left[\pi (1-cos^{2}\chi)sin^{2}\beta +2\pi cos^{2}\chi cos^{2}\beta \right]\\ \nonumber  
&\times&{f^{0}}^{2}(q)e^{E_{q}/T}\frac{\omega^{1/2}}{s}\\ \nonumber
&\times&\int_{-1}^{1}d(cos\hat{\theta})\frac{1}{g_{Q}}\sum \overline{|M|}^2\\ \nonumber
&\times&\int_{0}^{2\pi}F({\mathbf p}^{'})~d\hat{\phi}.
\label{thirtytwo}
\end{eqnarray}
Thus the total drag and/or diffusion coefficient for an anisotropic QGP is :
\begin{equation}
\langle\langle F(p)\rangle\rangle = \langle\langle F_{\rm{iso}}(p)\rangle\rangle + \langle\langle F_{\rm{aniso}}(p)\rangle\rangle.
\label{thirtythree}
\end{equation}

\section{Results and Discussions}
\subsection{For Isotropic QGP}
\begin{figure}
\includegraphics[scale=0.45]{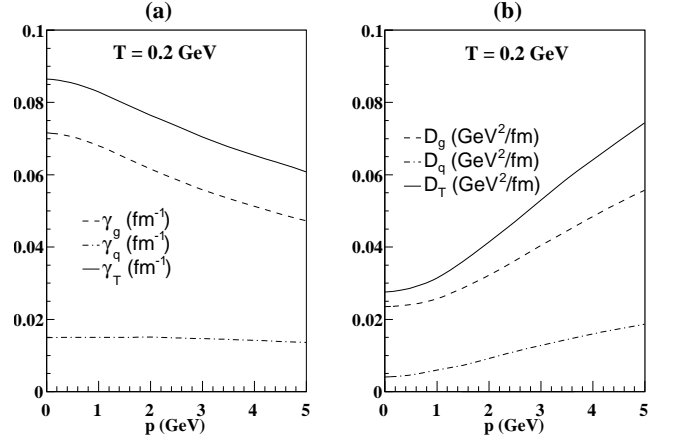}
\caption{(a) Variation of Drag Coefficient of charm quark 
with respect to initial momentum of heavy quark at a fixed 
QGP temperature $T=200$ MeV. Dash-dotted curve represents 
the drag on heavy quark due to light quarks of the QGP medium 
and dotted curve shows the contribution of gluons. Further 
solid curve is the sum of these two contributions. (b) 
Variation of diffusion coefficient of charm quark with 
respect to HQ momentum. All other things are similar to (a).}
\end{figure}

\begin{figure}
\includegraphics[scale=0.45]{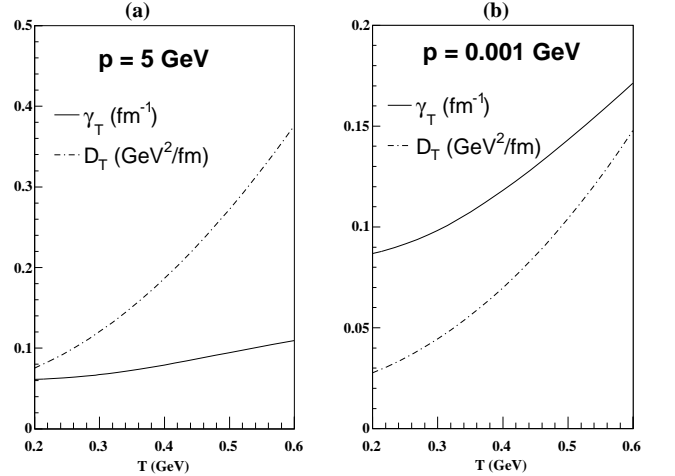}
\caption{Drag and diffusion coefficient with respect to temperature 
at a fixed HQ momentum (a) $p=5$ GeV, and (b) $p=0.001$ GeV.}

\end{figure}
First of all we would like to explore how the charm quarks 
are dragged by the partons while propagating in a hot and isotropic 
medium through its drag coefficient. To see the effects of intrinsic 
motion of test quark, we vary the momentum of charm quark 
from non-relativistic limit ($p=0.001$ GeV) to relativistic limit
($p$=5 GeV) in Figure 1(a) at some temperature, $T=0.2$ GeV of the
medium. 
During the evolution of heavy quarks in medium, the charm quarks are dragged 
by both quarks and gluons, thus we calculate separately the contribution by 
quarks, gluons and their sum total, which are shown by the 
dash-dotted, dotted and solid curves, respectively. 
In our calculation of the matrix element in $t$-channel, we take the 
regulator, $\mu$ as the leading-order Debye mass, 
$m_D(T)$, unlike a constant value used in other calculations. For the
temperature dependence of the Debye mass, we take the strong coupling ($g_s$) 
from one-loop expression. We found that the 
drag coefficient decreases when the momentum of the charm quark 
increases.
This observation agrees with common sense because the relative speed 
of the charm quark with respect to the medium
increases with the increase its momentum and hence the drag coefficient
decreases. The above observation can be understood from the point 
of view of statistical mechanics in the following way:
The equilibration rate of HQs in phase space decreases with the 
increase in its momentum, hence the drag coefficient for
HQs should decrease with its momentum because the drag coefficient is 
related linearly to the kinetic equilibration rate.
This understanding will later be useful 
to understand the variation of diffusion coefficient with the momentum 
(in Fig. 1(b)). Another observation of Fig. 1 (a) is that 
the momentum dependence of the drag coefficient are mostly emanated
from the gluon scattering due to their abundance and the large contribution
to the cross-section whereas the contribution by light quarks
is meagre. 

On the other hand the momentum dependence of the diffusion coefficient
is opposite, {\em i.e.} $D_T$
increases with the momentum (shown in Figure 1b) because it is easy 
for HQs having larger momentum to diffuse 
in the system compared to HQs of lower momentum. From the point
of view of statistical physics, the 
diffusion coefficient is a measure of the equilibration time 
(inverse of the equilibration rate) thus the coefficient should be
greater when the HQ momentum becomes larger. Like the drag coefficient, the 
gluons contribute
substantially to the momentum dependence of diffusion coefficient 
compared to the meagre contribution by quarks.

\begin{figure}
\includegraphics[scale=0.45]{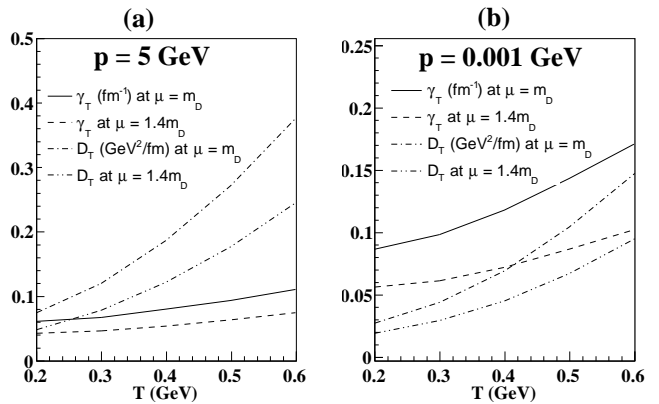}
\caption{(a) Drag and diffusion coefficients evaluated 
at fixed value of HQ momentum (a) $p=5$ GeV, and (b) $p=0.001$ GeV 
as function of temperature ($T$). We have plotted both coefficients 
at two different values of matrix regulator ie., $\mu=m_{D}$ and $1.4m_{D}$, 
where $m_{D}$ is the Debye mass.}
\end{figure}
In Fig. 2(a), we have studied how a relativistic charm quark is diffused
or dragged while evolving in a static isotropic QCD medium 
when the temperature of the medium changes from lower to higher values. 
We found that the drag coefficient becomes small and increases with
the temperature slowly whereas the diffusion coefficient increases with
the temperature rapidly, so their separation ($\gamma_T -D_T)$ increases
with the temperature. This can be understood qualitatively:
since the momentum is much higher than the temperature 
of the medium therefore the physical scale set here is the 
momentum of HQ only. 
If the temperature of a thermal medium is increased then the constituents
of the medium exert more and more random force on the test particle, 
thus the increase of temperature causes the motion of the test particle 
more random. Since the diffusion is asymptotically 
related to the temperature of the medium, therefore the diffusion
coefficient increase with the increase of the temperature.
Similar to Fig. 2a, in Figure 2 b, we have explored how do the drag 
and diffusion 
coefficients depend on the temperature for a non-relativistic
charm quark ($p=0.001$ GeV). Since the relative speed of 
HQ becomes small therefore the drag coefficient becomes large. By the same 
reasoning as in Figure 2 a, the relevant scale set here is the temperature 
of the medium, {\em not the momentum of the test HQ}. 
Thus both the equilibration rate and the random force exerted by the partons
increase with the temperature for HQs having low momentum. 
Hence both drag and diffusion coefficients increase with the temperature.
The increase of diffusion coefficient with the temperature 
in both figures (Figure 2 a \& b) is understandable because 
the number of constituents faced by the test particle increases 
with temperature (for massless case, $n\propto T^{3}$) and 
thus the random force exerted on HQ by the constituents increases. 

As we mentioned earlier in Figs 1 and 2, we have taken 
the regulator in the $t$ channel matrix element by the Debye mass 
in leading-order ($m_D^{LO}$). The Debye mass in the leading-order is 
correct in the weak coupling regime when the 
coupling constant is very small $g<<T$. However when $g\sim T$ then 
higher order corrections also arise in the 
Debye mass~\cite{vineet}. Kajantie et al.~\cite{kajantie} 
computed these contributions of 
$O(g^{2}T)$ and $O(g^{3}T)$ from a three-dimensional 
effective field theory. Here we wish to see the 
effect of the regulator on the drag and diffusion coefficients 
due to the corrections in the Debye mass.
Thus, we have used two regulators : $\mu=m_{D}^{LO}$ and $1.4m_{D}^{LO}$, 
where the factor $1.4$ takes into account the next-to-leading order 
corrections, for relativistic ($p$= 5 GeV) and non-relativistic
($p$=0.001 Gev) in Figures 3 a and b, respectively. 
We observed that the inclusion of higher order 
effects in the Debye mass as the regulator decreases both drag and 
diffusion coefficients. 
To be specific, the changes in drag coefficient of non-relativistic 
HQs due to increase in the regulator are about $35\% - 40\%$ while going 
from T=0.2 GeV to T=0.6 GeV whereas in relativistic regime the change in 
drag coefficient ($\approx 35\%$) is almost independent to the change 
in temperature.  Further the change in the value of diffusion is about $30\% -35\%$ as one goes from 
lower to higher temperature at non-relativistic momentum but again the percentage of change in diffusion 
 ($\approx 35\%$) remains independent to the temperature at relativistic momentum $p=5$ GeV. 
\\
\begin{figure}
\includegraphics[scale=0.40]{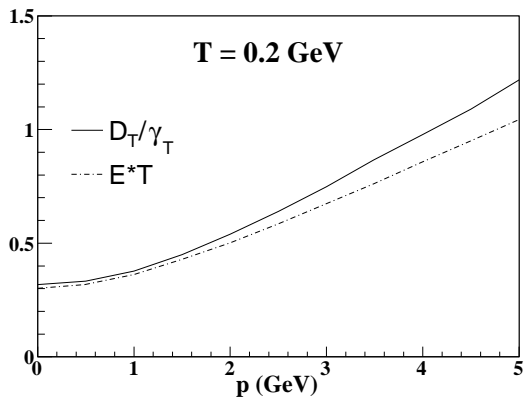}
\caption{Ratio of diffusion to drag ($D_{T}/\gamma_{T}$) 
with respect to HQ momentum $p$ 
at fixed value of QGP temperature $T=0.2$ GeV. 
We have also plotted the product of HQ energy ($E$) 
and temperature ($T$) by dash-dotted curve for comparison.}
\end{figure}

To check the validity of the fluctuation dissipation theorem (FDT) for a picture
where the non-equilibrated degrees of freedom (in this case it is heavy quarks)
evolves in the background of equilibrated degrees of freedom, we have 
studied the ratio of the diffusion to drag coefficient ($D_T/\gamma_T$) 
for dynamical HQs in isotropic medium at a temperature, T=0.2 GeV 
(in Figure 4). As we know that the ratio is asymptotically related to the 
temperature of the medium so we have also plotted the quantity 
${\rm{Energy (E)}}\times T$ 
in the same figure. We observed that the FDT is almost satisfied
in the non-relativistic limit of HQ momentum but is violated as the
HQ momentum becomes more and more relativistic. This observation 
seems more plausible because FDT is satisfied only if a non-equilibrated 
degrees of freedom evolves in an ideal heat bath and undergoes 
through linear damping~\cite{rafel}.
Our result is consistent with other calculations~\cite{dass}, where 
the KLN factorization is employed to model the pre-equilibrium 
momentum space gluon distribution~\cite{kln}. 
\begin{figure}
\includegraphics[scale=0.45]{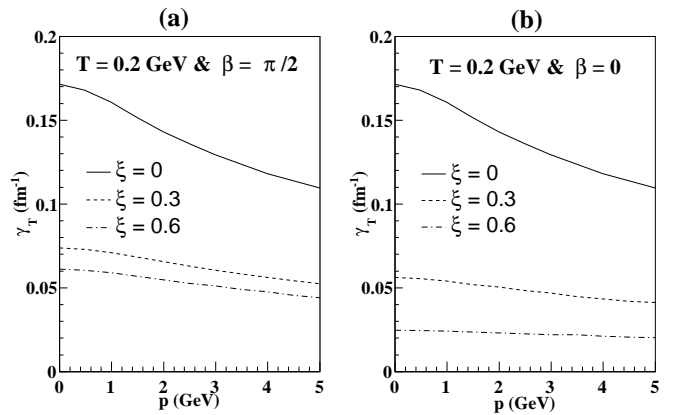}
\caption{Variation of drag coefficient with respect 
to HQ momentum for three different values of anisotropy parameter ie., $\xi=0,~0.3$, and $0.6$) at a medium temperature $T=0.2$ GeV for (a) perpedicular ($\beta=\pi/2$), and (b) parallel ($\beta = 0$) case.}
\end{figure}

\begin{figure}
\includegraphics[scale=0.45]{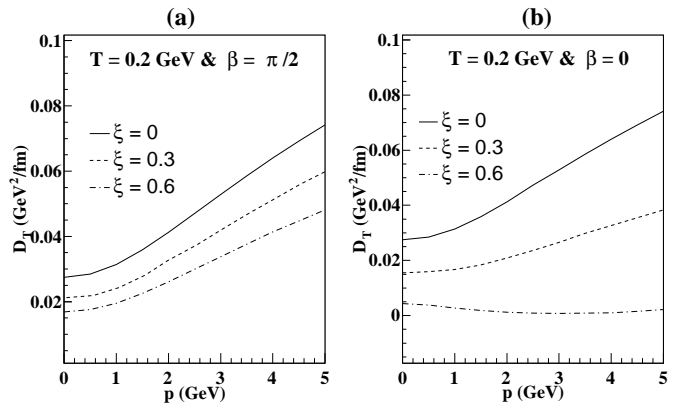}
\caption{Variation of diffusion coefficient with respect 
to HQ momentum for three different values of anisotropy parameter ie., $\xi=0,~0.3$, and $0.6$) at a medium temperature $T=0.2$ GeV for (a) perpedicular ($\beta=\pi/2$), and (b) parallel ($\beta = 0$) case.}
\end{figure}
\subsection{For anisotropic QGP}

As we discussed in the preamble that the system produced at the early
stage of ultra-relativistic heavy ion collisions exhibits an anisotropy 
in the momentum space, thus we aim to explore the effect of anisotropy on 
the heavy quark evolution because heavy quarks are also produced
at early stages of the collision. 
Since the anisotropy introduces an angular dependence in the drag 
coefficient so we have calculated the coefficient for different values
of anisotropy parameter for two cases: a) when the partons
move transverse to the direction of anisotropy ($\beta=\pi/2$) 
in Figure 5 a and 
b) when the partons moves along the direction of anisotropy 
($\beta=0$) in Figure 5 b.
The immediate observation is that the drag coefficient always decreases 
with the anisotropy ($\xi \ne 0$) for both parallel and perpendicular 
alignment, which can be understood qualitatively: 
In the small anisotropic limit, the anisotropic distribution function 
for partons may be approximated as an isotropic distribution function 
by removing particles with a large momentum component along the
direction of anisotropy, (${\bf n}$), which 
causes a reduction of the number of partons around a test heavy quark 
in a given phase-space point ($n_{\rm aniso}\approx n_{\rm iso}/\sqrt{1+\xi}$).
As a result, while propagating in the medium, HQ encountered less number of 
scatterings and hence the equilibration rate 
becomes smaller in anisotropic plasma, resulting a decrease in
the drag coefficient. It can also be seen that if the temperature of the
medium is increased then the (negative) correction  due to anisotropy 
increases, as a result the coefficient decreases sharply for all 
values of HQ momentum. Another important 
observation is that the drag coefficient of HQ in parallel alignment 
(${\mathbf q}||{\mathbf n}$) is always less than the 
perpendicular alignment (${\mathbf q}\perp {\mathbf n}$).
This is due to the fact that for parallel alignment the momentum of partons 
got effectively shifted towards higher momentum side 
($q^{2}\rightarrow q^{2}+\xi ({\mathbf q}\cdot {\mathbf n})^{2}_{||},
~\xi > 0$) thus a large chunk of higher momentum 
particles do not contribute to the scattering of partons with HQ whereas 
for perpendicular alignment, the shift of partons to higher momentum side 
does not arise that much.
Similarly we can see from Figs. 6 (a) and (b) that the diffusion 
coefficient also decreases due to the momentum space anisotropy.
Further, like drag coefficient, diffusion coefficient of HQ also becomes 
smaller in parallel alignment than the perpendicular alignment.
  


In conclusion, we have first revisited the propagation of
charm quarks in a hot isotropic medium of quarks and gluons 
by its transport coefficients - drag and diffusion coefficients 
and then explore the dependencies of the coefficients on the 
charm quark momentum, the temperature of the medium
and the regulator in the $t$-channel matrix element in the form of 
Debye mass. We have understood the results through the
connection of the coefficients with the equilibration rate
from the statistical mechanics point of view. With these understanding
from the isotropic medium we move on to a medium which exhibits a
momentum anisotropy and calculated both drag and diffusion coefficients
in an anisotropic medium.
The important finding of our calculation is that both drag and 
diffusion coefficients of charm quarks get waned due to the
momentum anisotropy, which may be created at the early stages of 
the relativistic collisions. 
Moreover the anisotropy causes a 
direction dependent sizable modifications to both drag and diffusion,
as a result the (negative) correction due to anisotropy, 
when the partons are moving parallel to the direction of anisotropy,
is larger than when the partons are moving perpendicular to the 
direction of anisotropy. 
Thus the study of drag and diffusion coefficients
in anisotropic medium will open up further applications to the
phenomenology of relativistic heavy ion collisions.
 
\noindent
\section{Acknowledgments}
PKS and BKP is thankful for financial assistance from Council of Scientific 
and Industrial Research (No. CSR-656-PHY), Government of India. Authors also 
acknowledge the fruitful discussion with S. Das, R. Rapp and H. van Hees 
during the course of this work.

\newpage
\end{document}